\documentclass{article}
\usepackage{amsmath,amssymb}
\newcommand{\doublespacing}{\let\CS=\@currsize\renewcommand{\baselinesstrech}
{2.0}\tiny\CS}
\begin{document}
\begin{center}
{\huge \bf Non-Hermitian Oscillator and $\cal{R}$- deformed Heisenberg Algebra}
\end{center}
\vspace{2cm}
\begin{center}
{\it R.Roychoudhury{\footnote {e-mail :
raj@isical.ac.in} and B. Roy{\footnote{e-mail : barnana@isical.ac.in}}\\
Physics \& Applied Mathematics Unit \\
Indian Statistical Institute \\
Kolkata 700108, India.}}\\
{\it P.P.Dube{\footnote{e-mail: ppdube1@gmail.com}\\
Garalgacha Surabala Vidyamandir\\
West Bengal, India}}
\end{center}
\vspace{1cm}
\begin{center}
{\bf Abstract}
\end{center}
A non-Hermitian generalized oscillator model, generally known as the Swanson model, has been studied in the framework of $\cal{R}$-deformed Heisenberg algebra. The non-Hermitian Hamiltonian is diagonalized by generalized Bogoliubov transformation. A set of deformed creation annihilation operators is introduced whose algebra shows that the transformed Hamiltonian has conformal symmetry. The spectrum is obtained using algebraic technique. The superconformal structure of the system is also worked out in detail. An anomaly related to the spectrum of the Hermitian counterpart of the non-Hermitian Hamiltonian with generalized ladder operators is shown to occur and is discussed in position dependent mass scenario.

\section {\bf Introduction}
Recently there has been a surge of interest in deformed Heisenberg algebra containing the reflection operator $\cal{R}$ (R-deformed Heisenberg algebra (RDHA) [1].
It appears in the context of Wigner quantization schemes [2], initiated in 1950 by Wigner [3] generalizing the bosonic commutation relations. The single mode algebra that appeared in the work of Wigner implicitly, was further generalized for the field systems, and led to the concept of parabosons and parafermions [4]. In this sense it has some properties of universality [5].
Hamiltonians with reflection operators have most notably arisen in the context of quantum many-body integrable systems of Calogero-Sutherland type [6]
and their generalizations with internal degrees of freedom [7]. In ref[8], a hidden nonlinear supersymmetry was revealed in purely parabosonic harmonic oscillator systems (and in related Calogero models with exchange interaction) with the help of RDHA, and a problem of the quantum anomaly related to nonlinear supersymmetry was identified. This was actually studied later and resulted in the discovery of the hidden supersymmetry in quantum mechanical systems with a local Hamiltonian [9]. The RDHA was applied for the universal description of bosons, fermions, anyons and supersymmetry in (2+1) dimensions. In particular, the non unitary finite and infinite dimensional representation of the RDHA algebra have been used to obtain extended supermultiplets of anyons, or of bosons and fermions[10]. Deformed Heisenberg algebra has also been used to construct $N = 2$ supersymmetric
quantum mechanics [11] and the associated exactly solvable models [12]. Symmetry algebras containing reflection operator in the representations of the
generators of the algebra have been examined in the context of quantum oscillators [13]. Also, mirror symmetry plays an important role in the design of
spin chains for quantum information transport [14]. Recently RDHA algebra has been employed for bosonization of supersymmetric quantum mechanics [15] and
for describing anyons in (2+1) [16] and (1+1) dimensions [17]. The findings of many interesting physical and mathematical properties associated
with these studies has prompted us to consider a non-Hermitian generalized oscillator model, widely known as the Swanson model [18] within the framework of RDHA algebra. Though many features of this particular Hamiltonian have been studied [19], to the best of our knowledge, this has not been reported so far in the literature.\\

The non-Hermitian generalized oscillator Hamiltonian is given by
\begin{equation}
\hat{H}_S = \omega (\hat{a}^{\dagger}\hat{a} + \frac{1}{2}) + \alpha \hat{a}^2 + \beta {\hat{a}^{\dagger 2}}, ~~~~~~~~~\omega,\alpha,\beta \in \mathbf{R}
\end{equation}
where $\hat{a}$, $\hat{a^{\dagger}}$ are bosonic harmonic oscillator annihilation and creation operators satisfying usual commutation relation $[\hat{a},\hat{a}^{\dagger}] = 1$. If $\alpha \neq \beta$, $H_S$ is non-Hermitian $\cal{PT}$-symmetric (or equivalently ${\cal{P}}$-pseudo-Hermitian, $\cal{P}$ being the parity operator) [20], possesses real spectrum given by $E_n = (n + \frac{1}{2})\Omega, n = 0, 1, 2, \cdots$ where $\Omega^2 = \omega^2 - 4 \alpha \beta$ so long as $\omega > \alpha + \beta$ and the eigenfunctions can be derived from those of the harmonic oscillator [18].\\
In the present work we shall study the Hamiltonian (1) with $a$, $a^{\dagger}$ satisfying the (anti)commutation relations of the RDHA algebra given by
\begin{equation}
\begin{array}{lcl}
[\hat{a}, {\hat{a}}^{\dagger}] &=& 1 + \nu\mathcal{R}\\
\{\mathcal{R}, \hat{a}\} &=& \{\mathcal{R}, {\hat{a}}^{\dagger}\} = 0\\
{\mathcal{R}}^2 &=& 1~~\mathcal{R}^{\dagger} = \mathcal{R}^{-1} = \mathcal{R}\\
\end{array}
\end{equation}
 where $\nu \in \mathbf {R}$ is a deformation parameter and $\mathcal{R}$ is the reflection operator, $\mathcal{R} = (-1)^{\mathcal{N}} = exp(i \pi \mathcal{N})$. The number operator [10]
\begin{equation}
\mathcal{N} = \frac{1}{2}\{\hat{a}^{\dagger}, \hat{a}\} - \frac{1}{2}(\nu + 1),~~~[\mathcal{N}, \hat{a}^{\dagger}] = \hat{a}^{\dagger}~~~[\mathcal{N}, \hat{a}] = -\hat{a}
\end{equation}
defines the Fock space
\begin{equation}
\mathcal{F} = \{|n>, n = 0, 1, 2, \cdots\},~~~ \mathcal{N}|n> = n|n>
\end{equation}
where $|n> = C_n(a^{\dagger})^n |0>,~~ n = 0, 1, \cdots ,~~ \hat{a}|0> = 0,~~ <0|0> = 1$,
\begin{equation}
C_n = ([n]_{\nu}!)^{-\frac{1}{2}},~~[0]_{\nu}! = 1,~~ [n]_{\nu}! = \prod_{i = 1}^n [l]_{\nu},~~ n\geq 1,~~ [l]_{\nu} = l +\frac{1}{2}(1 - (-1)^l)\nu
\end{equation}
For $\nu > -1$, the RDHA algebra has infinite dimensional, parabosonic type unitary irreducible representations; for $\nu = -(2r+1),~r = 1, 2, \cdots$ it has $(2r+1)$-dimensional, non unitary parafermionic-type representations; for $\nu < -1,~~\nu \neq -(2r+1)$, it has infinite dimensional non-unitary representations. In what follows we have not considered any particular representation for the $\mathcal{R}$ deformed Heisenberg algebra. But it is important to mention that if one takes especially the non-unitary finite and infinite dimensional representations as was done in [10] then this will have an effect on the generalized Bogoliubov transformation (6) as well as on the spectrum of ${\hat{H}}_S'$ discussed in section 4.

The objective of the present work is two-fold: i) First, to analyse the conformal and superconformal structure associated with the diagonalized version of the non-Hermitian Hamiltonian (1)  ii) Second, to obtain the equivalent Hermitian counterpart $\tilde{h}$ [21] of (1) (with generalized ladder operators) by a similarity transformation [22]
 and show how the generalized quantum rule (2) in this case gives rise to an anomaly related to the spectrum of $\tilde{h}$.  The reason for this apparent anomaly is attributed to the fact that the domain of $x$ and the domain of the variable used in the coordinate transformation to obtain the conformal Hamiltonian are not necessarily the same. Our findings will be illustrated in the framework of coordinate dependent mass models.

\section{\bf Non-Hermitian Oscillator Hamiltonian and Generalized Bogoliubov Transformation}
In this section we shall just outline the essential results as the derivation is practically the same as has been done by Swanson [18]. The two new operators $\hat{c}$, $\hat{d}$ are introduced by means of a generalized Bogoliubov transformation
\begin{equation}
\begin{array}{lcl}
\hat{c} &=& \displaystyle \frac{l_1}{(l_1 l_4 - l_2 l_3)} {\hat{a}}^{\dagger} - \frac{l_3}{(l_1 l_4 - l_2 l_3)} \hat{a}\\
\hat{d} &=& \displaystyle \frac{l_4}{(l_1 l_4 - l_2 l_3)} \hat{a} - \frac{l_2}{(l_1 l_4 - l_2 l_3)} {\hat{a}}^{\dagger} \\
\end{array}
\end{equation}
where the coefficients $l_j  = 1,2,3,4$ are taken to be complex numbers. It reduces to the standard Bogoliubov transformation when $l_4 = l_1^*$ and
$l_3 = l_2^*$. Equation (6)                                                                                                                           ) is written in matrix form as
\begin{equation}
\left(
  \begin{array}{c}
    \hat{d} \\
    \hat{c} \\
  \end{array}
\right)
= \frac{1}{(l_1 l_4 - l_2 l_3)}\left(
    \begin{array}{cc}
      l_4 & -l_2 \\
      -l_3 & l_1 \\
    \end{array}
  \right) \left(
            \begin{array}{c}
              \hat{a} \\
              \hat{a}^{\dagger} \\
            \end{array}
          \right)
 \end{equation}
Then the commutation relation (2) gives
 \begin{equation}
 [\hat{d},\hat{c}] = \frac{1}{(l_1 l_4 - l_2 l_3)}[\hat{a}, \hat{a}^{\dagger}] = \frac{1}{(l_1 l_4 - l_2 l_3)}(1 + \nu \mathcal{R})
\end{equation}
Inversion of the matrix in (7) gives
\begin{equation}
\begin{array}{lcl}
\hat{a} &=& l_1 \hat{d} + l_2 \hat{c}\\
\hat{a}^{\dagger} &=& l_3 \hat{d} + l_4 \hat{c}\\
\end{array}
\end{equation}
Substituting (9) into (1) yields
\begin{equation}
\hat{H}_S = \Lambda \hat{c} \hat{d} + \tilde{\alpha} \hat{c}^2 + \tilde{\beta} \hat{d}^2 + \epsilon + \frac{1}{2} \omega
\end{equation}
where
\begin{equation}
\begin{array}{lcl}
\Lambda &=& \omega(l_4 \l_1 + \l_3 \l_2) + 2 \alpha l_2 l_1 + 2 \beta l_4 l_3\\
\tilde{\alpha} &=& \omega l_2 l_4 + l_2^2 \alpha + l_4^2 \beta\\
\tilde{\beta} &=& \omega l_3 l_1 + l_1^2 \alpha + l_3^2 \beta\\
\epsilon &=& \frac{1}{(l_1 l_4 - l_2 l_3)}(1 + \nu \hat{R}) (\omega l_3 l_2 + \alpha l_1 l_2 + \beta l_3 l_4)\\
\end{array}
\end{equation}
The coefficients $l_j$ s are so chosen that
\begin{equation}
\tilde{\alpha} = \tilde{\beta} = 0
\end{equation}
are satisfied. If $\alpha = \beta = 0$ in (1), then (9) can be chosen to reduce to $\hat{c} = \frac{l_1}{l_1 l_4 - l_2 l_3}\hat{a}^{\dagger}$ and
$\hat{d} = \frac{l_4}{l_1 l_4 - l_2 l_3}\hat{a}$.
This yields the boundary conditions
\begin{equation}
\lim_{\alpha, \beta \rightarrow 0}\frac{l_{2,3}}{(l_1 l_4 - l_2 l_3)} = 0
\end{equation}
and
\begin{equation}
\lim_{\alpha, \beta \rightarrow 0} \frac{l_{1,4}}{\sqrt{(l_1 l_4 - l_2 l_3)}} = \lim_{\alpha, \beta \rightarrow 0}\frac{1}{(l_1 l_4 - l_2 l_3)}(1 + \nu \mathcal{R})
\end{equation}

Now the task is to find the solutions to (11) and (8) consistent with the limits (13) and (14). In the subsequent calculations, the value of $(l_1 l_4 - l_2 l_3)$ has been set to unity. Writing the constraints of (12) as two quadratic equations under the assumptions $l_1 \neq 0$ and $l_4 \neq 0$ and choosing appropriate signs consistent with the limits (13) and (14) while taking the square roots, gives [18]
\begin{equation}
\begin{array}{lcl}
l_1 l_4 &=& \displaystyle \frac{2 \alpha \beta}{(4 \alpha \beta - \omega^2 + \omega \sqrt{\omega^2 - 4 \alpha \beta})} \\
\l_2 l_3 &=& \displaystyle \frac{\omega - \sqrt{\omega^2 - 4 \alpha \beta}}{2\sqrt{\omega^2 - 4 \alpha \beta}} \\
l_1 \l_2 &=& - \displaystyle \frac{\beta}{\sqrt{\omega^2 - 4 \alpha \beta}} \\
l_3 l_4 &=& \displaystyle -\frac{\alpha}{\sqrt{\omega^2 - 4 \alpha \beta}}
\end{array}
\end{equation}
Therefore $\Lambda$ and $\epsilon$ in (11) are
\begin{equation}
\begin{array}{lcl}
\Lambda &=& \sqrt{\omega^2 - 4 \alpha \beta} \\
\epsilon &=& \frac{1}{2}(\Lambda - \omega)(1 + \nu \hat{R})\\
\end{array}
\end{equation}
It is worth noting that in (16) $\epsilon$ is a reflection dependent operator.
The transformed Hamiltonian (10) then reads
\begin{equation}
\hat{H}_S = \sqrt{\omega^2 - 4 \alpha \beta} \hat{c} \hat{d}  + \frac{1}{2}\sqrt{\omega^2 - 4 \alpha \beta}  (1 + \nu \hat{R}) - \frac{1}{2}\omega \nu \hat{R}
\end{equation}

In the coordinate representation [23], $\mathcal{R}$ is realized by the parity operator $\cal{P}$:
\begin{equation}
{\cal{P}} |x> = |-x> \Rightarrow \{{\cal{P}}, x\} = \{{\cal{P}}, p_x\} = 0 , ~~~~{{\cal{P}}}^{\dagger} = {{\cal{P}}}^{-1} = {\cal{P}}~~~{\cal{P}}^2 = 1 \label{e4}
\end{equation}
whereas the deformed ladder operators can be realized in the form [1,16]
\begin{equation}
\begin{array}{lcl}
{\hat{c}} &=& \frac{1}{\sqrt{2}}(-x + ip_x) \\
{\hat{d}} &=& \frac{1}{\sqrt{2}}(-x - ip_x))
\end{array}
\end{equation}
where
\begin{equation}
p_x = -i(\frac{d}{dx} - \frac{\nu}{2x} \cal{P})
\end{equation}
Substitution of (19) into (17) gives
\begin{equation}
{\hat{H}}_S' = \frac{1}{2} \{p^2 + x^2 + \frac{1}{4x^2}(\nu^2 - 2\nu\cal{P})\}
\end{equation}
where ${\hat{H}}_S' = \displaystyle \frac{{\hat{H}}_S - \frac{1}{2}\omega \nu \cal{P}}{\sqrt{\omega^2 - 4 \alpha \beta}},~~(\omega^2 \neq 4\alpha \beta)$


\section{\bf Conformal structure of ${\hat{H}}_S'$}
Let us recall that the conformal group $O(2,1)$ is spanned by three generators $H$, the Hamiltonian, $D$, the dilatation generator, and $K$, the conformal generator. These generators form together the algebra [24]
\begin{equation}
[H,D] = iH~~~~[K,D] = -iK~~~~[H,K] = 2iD
\end{equation}
If we define
\begin{equation}
R = \frac{1}{2}(K + H)~~~~~~~~~~~~~~~~~~~~S = \frac{1}{2}(K - H)
\end{equation}
then $R$, $D$ and $H$ satisfy the $O(2,1)$ algebra
\begin{equation}
[D,R] = iS~~~~~~~~~~[S,R] = -iD~~~~~~~~~~~~~~[S,D] = -iR
\end{equation}
Evidently, (23) corresponds to just a change of basis of the Lie algebra $O(2,1)$.
The subgroup generated by $R$ is compact (rotation group in a plane) while those generated by $S$ and $D$ are not, being of the boost type [24].
Let us take the following representations for $H$, $K$, $D$
\begin{equation}
\begin{array}{lcl}
H &=& \displaystyle \frac{1}{2} \{p^2 + \frac{\nu^2 - 2 \nu \cal{P}}{4 x^2}\}\\
K &=& \displaystyle \frac{x^2}{2}\\
D &=& \displaystyle -\frac{1}{4} (x p_x + p_x x)
\end{array}
\end{equation}
where $p = -i \frac{d}{dx}$ and $p_x$ is given by (20).
Then it is easy to verify that $2R = {\hat{H}}_S'$, $S$ and $D$ satisfy (24), i.e. ${\hat{H}}_S'$ has the conformal symmetry.
\section{\bf Spectrum of ${\hat{H}}_S'$}
Let us introduce the ladder operators
\begin{equation}
L_{\pm} = S \pm iD
\end{equation}
These play the role of ladder operators, since these, together with $R$, satisfy the commutation relations of $SL(2,R)$ algebra [24]
\begin{equation}
[R, L_{\pm}] = \pm L_{\pm}~~~~~~~~~~~~~~~~~~~~~~~[L_+, L_-] = -2R
\end{equation}
At this point it is worth noticing that $\hat{c}^2$, $\hat{d}^2$ and $\hat{c}\hat{d}$ given in equation (22), satisfy the following commutation relations
\begin{equation}
[\hat{c}^2, \hat{c}\hat{d}] = -2 \hat{c}^2~~~~~~~~[\hat{d}^2, \hat{c}\hat{d}] = 2\hat{d}^2~~~~~~~~~~[\hat{c}^2,\hat{d}^2] = -4\hat{c}\hat{d} - 2(1+ \nu \cal{P})
\end{equation}
These can be identified with the generators $L_-$, $L_+$ and $R$ of $SL(2,R)$ algebra in the following way
\begin{equation}
\hat{c}^2 = -2 L_+~~~~~~~~~~~\hat{d}^2 = -2 L_-~~~~~~~~~~~\hat{c}\hat{d} = 2R - \frac{(1+\nu\cal{P})}{2}
\end{equation}
In terms of $H$, $K$ and $D$ [24]
\begin{equation}
\begin{array}{lcl}
L_+ &=& \frac{1}{2}K-H+2iD)\\
L_- &=& \frac{1}{2}(K-H-2iD)\\
R &=& \frac{1}{2}(H+K)
\end{array}
\end{equation}
and the Casimir operator is given by

\begin{equation}
\begin{array}{lcl}
J^2 &=& R^2 + R - L_- L_+\\
&=&\displaystyle \frac{HK+KH}{2} - D^2\\
&=& \displaystyle \frac{g}{4} - \frac{3}{16}
\end{array}
\end{equation}
where
\begin{equation}
g = \frac{\nu^2 - 2 \nu {\mathcal{P}}}{4}
\end{equation}
Putting $J^2 = r_0 (r_0 - 1)$, one finds
\begin{equation}
\begin{array}{lcl}
r_0 &=& \frac{1}{2}(1 + \sqrt{g + \frac{1}{4}})\\
&=& \frac{1}{4}(\nu + 3) ~~~ \mbox {if}~~ {\cal{P}} = -1\\
&=& \frac{1}{4}(\nu + 1) ~~~ \mbox {if}~~ {\cal{P}} = +1
\end{array}
\end{equation}
The reason for taking the positive sign before the square root in (33) arises from the conditions to be satisfied by the wavefunction of the quantum mechanical system. In this case the required conditions are the vanishing of both the lowest eigenfunction and its first derivative as $x \rightarrow 0$ [24]. This requires $r_0 > \frac{3}{4}$ which is possible for the choice of positive sign.
Now we write the eigenvalue equation as
\begin{equation}
{\hat{H}}_S'|n> = \epsilon_n |n>
\end{equation}
where $|n>$ are the eigenstates and $|0>$ labels the vacuum state.\\
From the commutation relations (27), it follows that [24]
\begin{equation}
L_{\pm} |n, r_0> = C_{\pm} (n, r_0) |n \pm 1, r_0>
\end{equation}
implying that successive eigenvalues differ by unity. Using (31), one gets
\begin{equation}
\begin{array}{lcl}
|C_{\pm}(n, r_0)|^2 &=& \epsilon_n (\epsilon_n \pm 1) - J^2\\
&=&  \epsilon_n (\epsilon_n \pm 1) - r_0(r_0 -1) \geq 0
\end{array}
\end{equation}
so that~~ $\epsilon_n \geq r_0~,~\epsilon_n \leq -r_0$. Here we consider $\epsilon_n$ to be positive and hence $\epsilon_n > \frac{3}{4}$ (since $r_0 > \frac{3}{4}$), to obtain positive eigenvalues given by
\begin{equation}
\epsilon_n = r_0 + n~~,~~~~~n = 0,1,2, \cdots
\end{equation}
where $r_0$ is given by (33).

\section{\bf Superconformal structure associated with ${\hat{H}}_S'$}
To study the supersymmetric version of conformal quantum mechanics of section 3, we shall make use of the general construction by Witten [25]. The supersymmetric generalization of ${\hat{H}}_S'$ is [26]
\begin{equation}
\begin{array}{lcl}
\cal{H} &=& \frac{1}{2} \{ Q_c, Q_c^{\dagger}\}\\
&=& \frac{1}{2} \{p_x^2 + W^2 - BW^{\prime}\}
\end{array}
\end{equation}
where the grading operator $B = [\psi^{\dagger}, \psi] = \sigma_3$, $\sigma_3$ being the third component of Pauli spin matrices [34].
\begin{equation}
\begin{array}{lcl}
Q_c &=& (-i p_x + \frac{\sqrt{g}}{x}) \psi^{\dagger}\\
{Q_c}^{\dagger} &=& \psi(i p_x + \frac{\sqrt{g}}{x})\\
W(x) &=& \displaystyle \frac{-1 + \sqrt{1 + 2\nu^2 + 4(1+\nu^2)(g \pm \sqrt{g})}}{2x} - \displaystyle \frac{(\nu \pm 2\nu \sqrt{g})}{2x} ~ \mbox {for}~ B = \pm 1, \mathcal{P} = 1\\
&=& \displaystyle \frac{-1 + \sqrt{1 + 2\nu^2 + 4(1+\nu^2)(g \pm \sqrt{g})}}{2x} + \displaystyle \frac{(\nu \pm 2\nu \sqrt{g})}{2x} ~ \mbox {for}~ B = \pm 1, \mathcal{P} = -1
\end{array}
\end{equation}
In (38), $W(x)$ is the superpotential generating the conformal supersymmetric quantum mechanics and is obtained by solving a Riccati equation. For $W(x)$ to be real $g > 0$ giving rise to the constraint $(\nu \mp 1)^2 > 1$ (corresponding to $\mathcal{P} = \pm 1$) which is consistent with the condition $r_0 > \frac{3}{4}$ (see equation (33)). The spinor operators $\psi$ and $\psi^{\dagger}$ in (39) obey the anticommutation relation
\begin{equation}
\{\psi, \psi^{\dagger}\} = 1
\end{equation}

Introduction of an extra pair of spinor operators $S$ and $S^{\dagger}$ as
\begin{equation}
\begin{array}{lcl}
S^{\dagger} &=& \psi x\\
S &=& x \psi^{\dagger}\\
\end{array}
\end{equation}
endows the system with a richer algebraic structure, namely the superconformal algebra [1,27] which is given by
\begin{equation}
\begin{array}{lcl}
\frac{1}{2} \{ Q_c, Q_c^{\dagger}\} &=& \cal{H}\\
\frac{1}{2}\{S, S^{\dagger}\} &=& K\\
\frac{1}{2}\{Q_c, S^{\dagger}\} &=& -iD - \frac{1}{2}B(1+\nu {\cal{P}}) + \sqrt{g}\\
\frac{1}{2}\{Q_c^{\dagger}, S\} &=& +iD - \frac{1}{2}B(1+\nu {\cal{P}}) + \sqrt{g}\\
\end{array}
\end{equation}
All other anticommutators like $\{Q_c, Q_c\}, \{Q_c, S\}$ vanish.
Explicitly
\begin{equation}
\begin{array}{lcl}
\cal{H} &=& \displaystyle \frac{1}{2} (\mathbf{I} + p_x^2 + \frac{\mathbf{I} g + \sqrt{g} B (1-\nu {\cal{P}})}{x^2})\\
K &=& \displaystyle \frac{1}{2} x^2\\
D &=& \displaystyle \frac{1}{2} (p_x x + x p_x)\\
B &=& [\psi, \psi^{\dagger}]
\end{array}
\end{equation}

The fermionic raising and lowering operators are defined as
\begin{equation}
\begin{array}{lcl}
M &=& Q_c - S\\
M^{\dagger} &=& Q_c^{\dagger} - S^{\dagger}\\
N &=& Q_c + S\\
N^{\dagger} &=& Q_c^{\dagger} + S^{\dagger}
\end{array}
\end{equation}
which obey the anticommutator algebra
\begin{equation}
\begin{array}{lcl}
\frac{1}{2} \{M , M^{\dagger}\} &=& 2 {\cal{H}}_0 + \frac{1}{2}(1 + \nu {\cal{P}})B - \sqrt{g} \equiv {\cal{H}}_1\\
\frac{1}{2} \{N , N^{\dagger}\} &=& 2 {\cal{H}}_0 - \frac{1}{2}(1 + \nu {\cal{P}})B + \sqrt{g} \equiv {\cal{H}}_2 \\
-\frac{1}{4}\{M , N^{\dagger}\} &=& {\cal{L}}_-\\
-\frac{1}{4} \{M^{\dagger}, N\} &=& {\cal{L}}_+
\end{array}
\end{equation}
where \\
${\cal{H}}_0 = \frac{1}{2}({\cal{H}} + K)$ is the supersymmetric generalization of the generator of compact rotation $2R = \hat{H}_S'$,\\
each of ${\cal{H}}_1$ and ${\cal{H}}_2$ are supersymmetric Hamiltonians and\\
${\cal{L}}_{\pm}$ are the supersymmetric generalizations of the ladder operators defined in (30).\\
The superconformal algebra (42) closes since [33]

\begin{equation}
[B,N^{\dagger}] = - N^{\dagger},~~~[K, N^{\dagger}] = - S^{\dagger}
\end{equation}
\begin{equation}
[\mathcal{H}, N^{\dagger}] = -Q_c^{\dagger} = -N^{\dagger} + S^{\dagger}
\end{equation}
\begin{equation}
[B,S] = S, ~~~ [K,S] = 0
\end{equation}
\begin{equation}
[\mathcal{H},S] = Q_{c} = N - S
\end{equation}

In terms of the superpotentials $W_1(x)$ and $W_2(x)$ corresponding to ${\cal{H}}_1$ and ${\cal{H}}_2$ respectively, $M$, $M^{\dagger}$, $N$ and $N^{\dagger}$ are written as [26]
\begin{equation}
\begin{array}{lcl}
M &=& (-i p_x + W_1(x))\psi^{\dagger}\\
M^{\dagger} &=& \psi(i p_x + W_1(x))\\
N &=& \psi (i p_x + W_2(x))\\
N^{\dagger} &=& (-i p_x + W_2(x))\psi^{\dagger}
\end{array}
\end{equation}
The superpotentials $W_1(x)$, associated with ${\mathcal{H}_1}$ is given by [28]
\begin{equation}
W_1(x) = - \frac{u_0'(x)}{u_0(x)}
\end{equation}
where $u_0(x)$ (taking $\mathcal{P} = -1)$ is
\begin{equation}
u_0(x) = A_0 e^{-\frac{x^2}{2}} x^{\frac{1}{2}(\nu + 2 + 2\sqrt{g})} ~ _1F_1(1, \frac{(\nu + 3 + 2\sqrt{g})}{2}, x^2)
\end{equation}
$A_0$ being a constant, and $_1F_1 (a,b,z)$ is the confluent Hypergeometric function [29]. Similarly $W_2(x)$ can be obtained. The case $\mathcal{P} = +1$ can be treated analogously.


\section{\bf Reduction of the Hermitian counterpart of non-Hermitian generalized Oscillator into the conformal Hamiltonian $\hat{H}_S'$}
In this section our aim is to obtain the equivalent Hermitian counterpart of the Hamiltonian (1) (with generalized ladder operators) by a similarity transformation and study the associated spectrum. The reason for taking generalized ladder operators giving rise to a generalized quantum condition  (2) is that it allows access to those physical systems that are underlined by a coordinate dependent mass [30] whereby one can determine  the isospectrality/ nonisospectrality of the Hermitian Hamiltonian with the Hamiltonian $\mathcal{H}_S^{\prime}$.
The genralized ladder operators $\hat{a}, {\hat{a}}^{\dagger}$ are taken as
\begin{equation}
\hat {a} = A(x)\frac{d}{dx} + B(x) + f(x){\cal{P}}~~~~{\tilde{a}}^{\dagger} = -A(x)\frac{d}{dx} + B(x) - A^{\prime} - f(x)\cal{P}
\end{equation}
where $f(x)$ is a function to be determined later and prime denotes differentiation with respect to $x$. It is assumed that  $A(-x) = A(x), B(-x) = - B(x), f(-x) = -f(x)$. It is to be noted that both for $\mathcal{P} = +1$ and $-1$, $B(x)$ and $f(x)$ should have the same parity. In this case
\begin{equation}
[\hat{a}, \hat{a}^{\dagger}] = 2AB' -4Bf{\cal{P}} + 2A^{\prime}f{\cal{P}} -A A''
\end{equation}
Substitution of (53) into (1) and removing the first derivative term of the resulting Hamiltonian with the help of a similarity transformation
gives the equivalent Hermitian Hamiltonian $\tilde{h}$ [21] and is given by
\begin{equation}
\tilde{h} = \tilde{{\rho}}_{(\alpha, \beta)} {H_S} \tilde{{\rho}}_{(\alpha, \beta)}^{-1} = -\frac{d}{dx} A^2 \frac{d}{dx} + V_{eff}(x)
\end{equation}
where
\begin{equation}
\tilde{\rho}(\alpha, \beta) = A(x)^{\frac{\alpha - \beta}{2}} exp(-(\alpha - \beta) \int^x\frac{B(x')}{A(x')}dx')
\end{equation}
\begin{equation}
\begin{array}{lcl}
V_{eff}(x) &=& \frac{1}{2}(\alpha + \beta)A A'' + [\frac{\alpha + \beta}{2} + \frac{(\alpha - \beta)^2}{4}]A'^2 - [1 + 2(\alpha + \beta) + 2(\alpha - \beta)^2]A'B\\
&& +[1 + 2(\alpha + \beta) + (\alpha - \beta)^2]B^2 - (\alpha + \beta + 1)AB' + \frac{1}{2}(\alpha + \beta +1) + f^2 \\
&& \mp Af' \mp (\alpha + \beta +1)A'f \pm  2(\alpha + \beta + 1) f B ~~~~\mbox{for} \mathcal{P} = \pm 1\\
\end{array}
\end{equation}
taking $\omega - \alpha - \beta = 1$.
It is important to mention here that $\tilde{\rho}(\alpha, \beta)$ should be well defined on $\mathbf {R}$ [21] so that the eigenfunctions of $\tilde{h}$ are normalizable. In fact this is consistent with the results obtained in section 6.1. There is a one to one correspondence between the energy eigenvalues of $\tilde{h}$ and $H_S$. Also, if $\psi_n(x)$ are the wave functions of the equivalent Hermitian Hamiltonian $\tilde{h}$ then the wave functions of the Hamiltonian $H_S$ are given by $(\tilde{\rho}_{(\alpha, \beta)})^{-1}\psi_n(x)$.

If $\hat{a}$ and ${\hat{a}}^{\dagger}$ satisfy the generalized quantum rule
\begin{equation}
[\hat{a}, {\hat{a}}^{\dagger}] = 1 + \nu \cal{P}
\end{equation}
then comparison of (54) with (58) gives
\begin{equation}
\begin{array}{lcl}
z(x) &=& \displaystyle \int^x \frac{dx'}{A(x')}\\
B(x) &=& \displaystyle -\frac{z''}{2z'^2} + \frac{z}{2}\\
f(x) &=& \displaystyle \frac{k_1}{z}\\
\end{array}
\end{equation}
where $k_1 = - \frac{\nu}{2}$ and for simplicity the integration constant is taken to be zero. Consequently, the Hamiltonian $\tilde{h}$ given in equation (55) becomes
\begin{equation}
\begin{array}{lcl}
\tilde{h} &=& -\frac{d}{dx} A^2 \frac{d}{dx} + V_{eff}(x)\\
V_{eff}(x) &=& \frac{z'''}{2z'^3} - \frac{5}{4}\frac{z''^2}{z'^4} + {\tilde{\omega}}^2 z^2 \pm (1+\alpha+\beta)k_1 + \frac{k_1(k_1\pm 1)}{z^2}~~~\mbox{for} \mathcal{P} = \pm 1
\end{array}
\end{equation}
where $ {\tilde{\omega}}^2 = \frac{(\alpha - \beta)^2 + 2(\alpha + \beta) + 1}{4}$.
For the change of variable (59) we have $A' = \frac{\dot{A}}{A}$,~~$A'' = \frac{\ddot{A}}{A^2} - \frac{\dot{A}^2}{A^3}$ where 'dot' denotes derivative with respect to $z$. Consequently the eigenvalue equation for the Hamiltonian $\tilde{h}$ given in (60), reduces to
\begin{equation}
-\frac{d^2\phi}{dz^2} + [\tilde{\omega}^2 z^2 + \frac{\nu(\nu + 2)}{4z^2}] \phi(z) = [E - \frac{(1+\alpha + \beta)\nu}{2}]\phi(z)
\end{equation}
where $\mathcal{P}$ has been taken to be equal to $-1$. Though in the subsequent calculations we have taken $\mathcal{P} = -1$, similar analysis can be made for $\mathcal{P} = 1$ also.
It is easily seen that equation (61) is the Schr$\ddot{o}$dinger equation for the Hamiltonian $\mathcal{H}_S'$ given in (21) for $\mathcal{P} = -1$. Equation (61) can be transformed into the Kummer's equation [29]
\begin{equation}
y\frac{d^2\chi}{dy^2} + \frac{d\chi}{dy}[\frac{(1-\nu)}{2} - y] + \chi(y)[\frac{E'}{4\tilde{\omega}} - \frac{1}{4} + \frac{\nu}{4}] = 0
\end{equation}
where $E' = E - \frac{(1 + \alpha + \beta)\nu}{2}$, by the transformations
\begin{equation}
y = \tilde{\omega} z^2~~~~~~~~~~~~~~~\phi(y) = y^{-\frac{c_1}{4}} e^{-\frac{y}{2}} \chi(y)
\end{equation}
Therefore the general solutions of equation (61) are given by
\begin{equation}
\begin{array}{lcl}
\phi_e(z) = N_e (\tilde{\omega}z^2)^{\frac{1}{2}+ \frac{\nu}{4}} e^{-\frac{\tilde{\omega} z^2}{2}} _1{F}_1 (\frac{1-\nu}{4} - \frac{E'}{4\tilde{\omega}}, \frac{1-\nu}{2}, \tilde{\omega} z^2)\\
\phi_o(z) = N_o (\tilde{\omega} z^2)^{\frac{1}{2} + \frac{\nu}{4}} e^{-\frac{\tilde{\omega} z^2}{2}} _1{F}_1 (\frac{\nu+3}{4} - \frac{E'}{4\tilde{\omega}}, \frac{\nu+3}{2}, \tilde{\omega} z^2)\\
\end{array}
\end{equation}

where $\phi_e$ and $\phi_o$ denote respectively even and odd parity solutions and $N_e$, $N_o$ are normalisation constants. The eigenfunctions of the Hamiltonian $\tilde{h}$ are given by
\begin{equation}
\psi(x) \sim A(z)^{-\frac{1}{2}} \phi(z)
\end{equation}
where $z(x)$ is given by equation (59).
\subsection{Breaking of Isospectrality}
In what follows we shall take the odd parity solution in (64) corresponding to $\mathcal{P} = -1$. Similar results can be obtained for $\mathcal{P} = 1$ as well. To obtain the eigenvalues of equation (61) one has to analyze the behaviour of the odd parity eigenfunction in (64). If the domain of the argument $\tilde{\omega}{z^2}$ is unbounded then $\phi_o(z)$ will not in general, square integrable because of the asymptotic behaviour of the confluent Hypergeometric function [29],
\begin{equation}
_1{F}_1(a,b,y) = \frac{\Gamma(b)}{\Gamma(a)} e^y y^{a-b}[1 + O(y^{-1})]
\end{equation}
Therefore to make the eigenfunctions square integrable one must take $a = -m (m = 0,1,2,3 \cdots)$ in which case $_1{F}_1(a,b,y)$ reduces to a polynomial. Correspondingly
\begin{equation}
E_{2m + 1} = \frac{(1 + \alpha + \beta) \nu}{2} + 2\tilde{\omega} (2m + \frac{(\nu+3)}{2})
\end{equation}
Hence in this case the Hamiltonian corresponding to the eigenvalue equation (61) has the spectrum of the Hamiltonian $\mathcal{H}_S'$ subject to the constant $\frac{(1 + \alpha + \beta) \nu}{2}$\\
But if the domain of $\tilde{\omega}z^2$ is finite then the odd parity eigenfunctions in (64) must vanish at the end points of the domain of $z$ and the eigenvalues will be given by the zeroes of the confluent hypergeometric functions when the arguments attain the end points. The first approximation of the m'th $(m = 1,2 \cdots)$ positive zero $X_0$ of $_1{F}_1(a,b,y)$ is given by [29]
\begin{equation}
X_0 = \frac{\pi^2(m + \frac{b}{2} - \frac{3}{4})^2}{2b - 4a}[1 + O(\frac{1}{\frac{b}{2} - a)^2})], ~~~m = 1,2, \cdots
\end{equation}
Hence from equation (64) we obtain, to the leading order (i.e. when $\frac{1}{(\frac{b}{2} - a)^2}$ is small in (68) which corresponds to large $\frac{E'}{4 \tilde{\omega}}$)
\begin{equation}
E_n = \frac{(1 + \alpha + \beta) \nu}{2} + \frac{\tilde{\omega}\pi^2}{4{z_\pm}^2}[n+1+\frac{\nu}{2}]^2~~~~n = 1, 3, 5, \cdots
\end{equation}
where $z_\pm$ are the end points of the argument of the confluent hypergeometric function. Hence in this the Hamiltonian corresponding to the eigenvalue equation (61) is not isospectral to the Hamiltonian $\mathcal{H}_S'$.
\section{Connection with coordinate dependent mass models}
As mentioned earlier the Swanson Hamiltonian with generalized ladder operators enables one to connect it to those physical systems which are endowed with coordinate dependent mass by choosing $A(x) = m(x)^{-\frac{1}{2}}$ which is a strictly positive function [22]. In this case the Hamiltonian (55) reduces to the coordinate dependent mass Hamiltonians [30] for which the corresponding Schr$\ddot{o}$dinger equation reads
\begin{equation}
(-\frac{d}{dx}\frac{1}{m(x)}\frac{d}{dx} + V_{eff}(x))\psi(x) = E \psi(x)
\end{equation}
Example 1. Unbroken Isospectrality \\
Let us consider the mass function $m(x) = e^x$. This mass function is used in studying the transport properties of semiconductors [31]. Correspondingly
$z(x) = 2 e^{\frac{x}{2}}$ which belongs to $(0, \infty)$ as $x \in (-\infty, \infty)$. This choice of mass function gives rise to the effective potential
\begin{equation}
V_{eff}(x) = - \frac{e^{-x}}{16} [4k_1^2 + 4k_1 -3] + 4 e^x \tilde{\omega} - k_1(1 + \alpha + \beta)
\end{equation}
 In this case, the eigenvalues of the Hamiltonian corresponding to the effective potential (71) is given by (67) i.e. the Hamiltonian is isospectral to the Hamiltonian (21)with $\mathcal {P} = -1$.\\

Example 2. Broken Isospectrality\\
In this case the mass function $m(x) = sech^2(x)$ which depicts the solitonic profile [32] is taken. Correspondingly $z(x) = \tan^{-1}(\sinh x)$. Clearly $z(x) \rightarrow z_\pm (=\pm \frac{\pi}{2})$ as $x \rightarrow \pm \infty$. The effective potential is
\begin{equation}
V_{eff}(x) = \frac{1}{4} - \frac{3}{4}\cosh^2 x + \tilde{\omega}^2 [\tan^{-1}(\sinh x)]^2 + \frac{k_1(k_1 + 1)}{[\tan^{-1}(\sinh x)]^2} - (1 + \alpha + \beta) k_1
\end{equation}
The eigenvalues of the Hamiltonian $\tilde{h}$ corresponding to the effective potential (72) is given by (69) putting $z_{\pm} = \pm \frac{\pi}{2}$
\begin{equation}
E_n = \frac{(1 + \alpha + \beta) \nu}{2} + \tilde{\omega}[n+1+\frac{\nu}{2}]^2~~~~n = 1, 3, 5, \cdots
\end{equation}
Therefore in this case the Hamiltonian is nonisospectral to the Hamiltonian (21) with $\mathcal {P} = -1$.
\section{Conclusion}
A generalized Bogoliubov transformation is used to diagonalize the non-Hermitian oscillator Hamiltonian. Within the framework of $\cal{R}$ deformed Heisenberg algebra the diagonalized Hamiltonian is shown to possess the conformal symmetry. The supersymmetric generalization of the above Hamiltonian and the associated superpotential have been constructed. Two Hamiltonians constructed by the fermionic raising and lowering operators are found to be supersymmetric in nature and the superpotentials associated with each of these are worked out. The Hermitian counterpart of the non-Hermitian oscillator Hamiltonian has been studied by using a generalized form of the ladder operators but keeping the generalized quantum rule (2) intact. This reveals an intriguing result in the sense that in this case the resulting Hamiltonian does not always possess the spectrum of the conformal Hamiltonian. The reason for this apparent anomaly is due to the peculiarities of the transformation involved. Our findings has been explained in the physically realistic quantum mechanical models with coordinate dependent mass.

\section{\bf Reference}
\begin{enumerate}
\item[1.] L.Brink, T.H.Hansson and M.A.Vasilev, Phys.Letts. {\bf B311} (1992) 109\\
    L.Brink, T.H.Hansson, S.Konstein and M.V.Vasilev, Nucl. Phys. {\bf B401} (1993) 591\\
    M.V.Vasilev, Pis'ma JETP {\bf 50} (1989) 344\
\item[2.] V.I.Manko, G.Marmo, E.C.G.Sudarshan and F.Zaccaria, Int.Jour.Mod.PHys. {\bf B11} (1997) 1296\\
          A.Horzela, Turk.Jour.Phys. {\bf 23} (1999) 903\\
          A.Horzela, Czech.Jour.Phys. {\bf 50} (2000) 1245\\
          E.Kpuscik, Czech.Jour.Phys. {\bf 50} (2000) 1279\
\item[3.] E.P.Wigner, Phys.Rev. {\bf 77} (1950) 711\
\item[4.] H.S.Green, Phys.Rev. {\bf 90} (1953) 270\\
          D.V.Volkov, Sov.Phys.JETP {\bf 9} (1959) 1107\\
          O.W.Greenberg, Phys.Rev.Letts. {\bf 13} (1964) 598\\
          O.W.Greenberg and A.M.L.Messiah, Phys.Rev. {\bf B138} (1965) 1155\\
          A.B.Govorkov, Theor.Math.Phys. {\bf 54} (1983) 234\
\item[5.] M.S.Plyushchay, Nucl.Phys. {\bf B491} (1997) 619\
\item[6.] F.Calogero, Jour.Math.Phys. {\bf 10} (1969) 2191\\
          B.Sutherland, Jour.Math.Phys. {\bf 12} (1971) 246\\
          A.P.Polychronakos, Phys.Rev.Letts. {\bf 69} (1992) 703\\
          L.Lapointe and L.Vinet, Commun.Math.Phys. {\bf 178} (1996) 425\\
          R.Floreanini, L.Lapointe and L.Vinet, Phys.Letts. {\bf B389} (1996) 327\
\item[7.] J.A.Minahan and A.P.Polychronakos, Phys.Letts. {\bf B302} (1993) 265\
\item[8.] M.S.Plyushchay, Int.Jour.Mod.Phys.{\bf A 15} (2000) 3679\
\item[9.] F.Correa and M.S.Plyushchay, Annal.Phys. {\bf 322} (2007) 2493\\
          F.Correa and M.S.Plyushchay, Annal. Phys. {\bf 327} (2012) 1761\
\item[10.] P.A.Horvathy, M.S.Plyushchay, M.Valenzuela, Annal.Phys. {\bf 325} (2010) 1931\
\item[11.] M.S.Plyushchay, arXive:hep-th/9404081\
\item[12.] S.Post, L.Vinet and A.Zhedanov, arXiv:math-ph/1107.5844v2\
\item[13.] E.I.Jafarov, N.I.Stoilova and J.Van der jeugt, SIGMA {\bf 8} (2012) 025 and the references cited therein\
\item[14.] A.Kay, Int.Jour.Quant.Inf. {\bf 8} (2010) 641\
\item[15.] T.Brzezinski, I.L.Egusquiza and A.J.Macfarlane, Phys.Letts. {\bf B311} (1993) 202\\
           M.Plyushchay, Mod.Phys.Letts. {\bf A11} (1996) 2953\\
           M.S.Plyushchay, Annal.Phys. {\bf 245} (1996) 339\
\item[16.] M.S.Plyushchay, Phys.Letts. {\bf 320} (1994) 91\
\item[17.] U.Aglietti, L.Griguolo, R.Jackiw, S-Y. Pi and D.Seminara, arXiv:hep-th/9606141\
\item[18.] M.S.Swanson, Jour.Math.Phys. {\bf 45} (2004) 585\
\item[19.] H.F.Jones, Jour.Phys. {\bf A38} (2005) 1741\\
           D.P.Musumbu, H.B.Geyer and W.D.Heiss, Jour.Phys. {\bf A40} (2007) F75\\
           C.Quesne, Jour.Phys. {\bf A40} (2007) F745\\
           B.Bagchi and T.Tanaka, Phys.Letts. {\bf A 372} (2008) 5390\\
           A.Sinha and P.Roy, Jour.Phys. {\bf A40} (2007) 10599\\
           A.Sinha and P.Roy, Jour.Phys. {\bf A 42} (2009) 052002\\
           B.Midya, P.P.Dube and R.Roychoudhury, Jour.Phys. {\bf A44} (2011) 062001 (FTC)\
\item[20.] C.M.Bender, Rep.Prog.Phys. {\bf 70} (2007) 947\\
           A.Mostafazadeh, Phys.Scrip. {\bf 82}(2010) 038110\
\item[21.] F.G.Scholtz, H.B.Geyer and J.Hahne, Ann.Phys. {\bf 213} (1992) 74\\
           A.Mostafazadeh and J.Batal, Jour.Phys. {\bf A37} (2004) 11645\\
           R.Kretschmer and L.Szymanowski, Phys.Letts. {\bf A325} (2004) 112\
\item[22.] B.Bagchi, C.Quesne, R.Roychoudhury, Jour.Phys. {\bf A38} (2005) L647\
\item[23.] L.M.Yang, Phys.Rev. {\bf 84} (1951) 788\\
           H.L.Carrion and R.De Lima Rodrigues, Mod.Phys.Letts. {\bf A 25} (2010) 2507\
\item[24.] V.de Alfaro, S.Fubini and G.Furlan, Nuovo Cimento {\bf A34} (1976) 569\
\item[25.] E.Witten, Nucl.Phys.B {\bf 185} (1981) 513\\
           E.Witten, Nucl.Phys.B {\bf 202} (1982) 253\
\item[26.] S.Fubini and E.Rabinovici, Nucl.Phys.B {\bf 245} (1984) 17\
\item[27.] R.Haag, V.T.Lopuszanski, M.Sohnius, Nucl.Phys.B {88} (1975) 257\\
           W.Nahm, V.Rittenberg, M.Scheunet, Phys.Letts. {\bf B61} (1976) 383\
\item[28.] F.Cannata, G.Junker, J.Trost, Phys.Letts. {\bf A 246} (1998) 219\
\item[29.] M.Abramowitz and I.A.Stegun, Handbook of Mathematical Functions (1970) (Dover, New York)\
\item[30.] J.BenDaniel and C.B.Duke, Phys.Rev.{\bf B152} (1966) 683\
\item[31.] R.Koc, G.Sahinoglu and M.Koca, Eur.Phys.Jour. {\bf B 48} (2005) 583\
\item[32.] B.Bagchi, Jour.Phys. {\bf A40} (2007) F1041\
\item[33.] R.P.Martinez-y-Romero and A.L.Salas-Brito, Jour.Math.Phys. {\bf 33} (1992) 1831\
\item[34.] F.Cooper, A.Khare and U.Sukhatme, Supersymmetry and Quantum Mechanics (2001), World Scientific, Singapore.
\end{enumerate}
\end{document}